\documentclass[aps,prb,floatfix,preprint,nofootinbib]{revtex4}
\usepackage{subfigure}
\usepackage{amssymb,amsmath}
\usepackage{graphicx}
\usepackage{color}
\usepackage{rotating}
\usepackage{booktabs}
\newcommand{\bra}[1]{\mbox{$\langle #1|$}}
\newcommand{\ket}[1]{\mbox{$|#1\rangle$}}

\long\def\symbolfootnote[#1]#2{\begingroup%
\def\thefootnote{\fnsymbol{footnote}}\footnote[#1]{#2}\endgroup}

\usepackage{hyperref}

\newcommand{\ad}{\hat{a}^+}

\newcommand{\I}{\mathbf{1}}

\renewcommand{\d}{\mathrm{d}}

\begin{document}

\title{Towards Quantum Chemistry on a Quantum Computer}

\author{B.~P.~Lanyon$^{1,2}$, J.~D.~Whitfield$^{4}$, G.~G.~Gillett$^{1,2}$, M.~E.~Goggin$^{1,5}$, M.~P.~Almeida$^{1,2}$, \\
I.~Kassal$^{4}$, J.~D.~Biamonte$^{4,}$\footnote{Present address: Oxford
  University Computing Laboratory, Oxford OX1 3QD, United Kingdom.}, M.~Mohseni$^{4}$,
B.~J.~Powell$^{1,3}$, M.~Barbieri$^{1,2,}$\footnote{Present address: Laboratoire Ch. Fabry de l'Institut d'Optique, Palaiseau, France.},
A.~Aspuru-Guzik$^{4}$ \& A.~G.~White$^{1,2}$}
\affiliation{$^{1}$Department of Physics, $^{2}$Centre for Quantum Computer Technology, $^{3}$Centre for Organic Photonics \& Electronics, University of Queensland, Brisbane 4072, Australia\\
$^{4}$Department of Chemistry and Chemical Biology, Harvard University, Cambridge, MA 02138, USA\\
$^{5}$ Department of Physics, Truman State University, Kirksville, MO 63501, USA}

\maketitle

{\bf \noindent The fundamental problem faced in quantum chemistry is the calculation of molecular
properties, which are of practical importance in fields ranging from materials science to biochemistry. Within chemical precision, the total energy of a molecule as well as most other properties, can be calculated by solving the Schr\"odinger equation. However, the computational resources required to obtain exact solutions on a conventional computer generally increase exponentially with the number of atoms involved~\cite{Fey82,Llo96}. This renders such calculations intractable for all but the smallest of systems.  
Recently, an efficient algorithm has been proposed enabling a quantum computer to overcome this problem by achieving only a polynomial resource scaling with system size~\cite{Llo96, Abrams1997, zalka-1998-454}. Such a tool would therefore provide an extremely powerful tool for new science and technology. Here we present a photonic implementation for the smallest problem: obtaining the energies of H${_2}$, the hydrogen molecule in a minimal basis. We perform a key algorithmic step---the iterative phase estimation algorithm~\cite{LidarWang99, AGDL+05, BCC06, CBMG08}---in full, achieving a high level of precision and robustness to error. We implement other algorithmic steps with assistance from a classical computer and explain how this non-scalable approach could be avoided. Finally, we provide new theoretical results which lay the foundations for the next generation of simulation experiments using quantum computers. We have made early experimental progress towards the long-term goal of exploiting quantum information to speed up quantum chemistry calculations. }

Experimentalists are just  beginning to command the level of control over quantum systems required  to explore their information processing capabilities. An important long-term application is to simulate and calculate properties of other many-body quantum systems. Pioneering experiments were first performed using nuclear-magnetic-resonance--based systems to simulate quantum oscillators~\cite{STHL+99}, leading up to recent simulations of a pairing Hamiltonian~\cite{YWXD06,BCC06}.  Very recently the phase transitions of a two-spin quantum magnet were simulated~\cite{Friedenauer:2008qm} using an ion-trap system. Here we simulate a quantum chemical system \emph{and calculate its energy spectrum}, using a photonic system. 

Molecular energies are represented as the eigenvalues of an associated time-independent Hamiltonian $\hat{H}$ and can be efficiently obtained to fixed accuracy, using a quantum algorithm with three distinct steps~\cite{AGDL+05}: encoding a molecular wavefunction into qubits; simulating its time evolution using quantum logic gates; and extracting the approximate energy using the phase estimation algorithm~\cite{Kit95,Abrams1997}. The latter is a general-purpose quantum algorithm for evaluating the eigenvalues of arbitrary Hermitian or unitary operators. The algorithm  estimates the phase, $\phi$, accumulated by a molecular eigenstate, $\ket{\Psi}$, under the action of the time-evolution operator, $\hat{U}{=}e^{-i\hat{H}t/\hbar}$, i.e.,

\begin{equation}
\label{eq:readout}
e^{-i\hat{H}t/\hbar}\ket{\Psi}{=}e^{-iEt/\hbar}\ket{\Psi}{=}e^{-i2\pi\phi}\ket{\Psi}
\end{equation}
where $E$ is the energy eigenvalue of $\ket{\Psi}$. Therefore, estimating the phase for each eigenstate amounts to estimating the eigenvalues of the Hamiltonian. 

We take the standard approach to quantum-chemical calculations by solving  an approximate Hamiltonian created by employing the Born-Oppenheimer approximation (where the electronic Hamiltonian is parameterized by the nuclear configuration) and choosing a suitable truncated basis set in which to describe the non-relativistic electronic system. Typical sets consist of a finite number of single-electron atomic orbitals, which are combined to form antisymmetric multi-electron molecular states (configurations)~\cite{Helgaker2000}. Calculating the eigenvalues of the electronic Hamiltonian using all configurations gives the exact energy in the basis set and is referred to as full configuration interaction (FCI). For $N$ orbitals and $m$ electrons there are ${N \choose m}{\approx} N^{m}/m!$ ways to allocate the electrons among the orbitals.  This exponential growth is the handicap of FCI calculations on classical computers.

As described in the Methods Summary, the Hamiltonian is block diagonal in our choice of basis, with 2x2 sub-matrices ($\hat{H}^{(1,6)}$ and $\hat{H}^{(3,4)}$). We map the configurations spanning each sub-space to the qubit computational basis. Since the subspaces are two-dimensional, one qubit suffices to represent the wavefunction. The corresponding time-evolution operators, $\hat{U}^{(p,q)}{=}e^{-i\hat{H}^{(p,q)}t/\hbar}$---where $(p,q){=}(1,6)$ or $(3,4)$---are therefore one-qubit operators. Finding the eigenvalues of each separately, using a phase estimation algorithm, amounts to performing FCI. 
For the purpose of our demonstration, we encode exact eigenstates, obtained via a preliminary calculation on a classical computer. In our Appendix we show that the algorithm is in fact robust to imperfect eigenstate encoding.

\begin{figure}[p]
\includegraphics[width=0.8 \columnwidth]{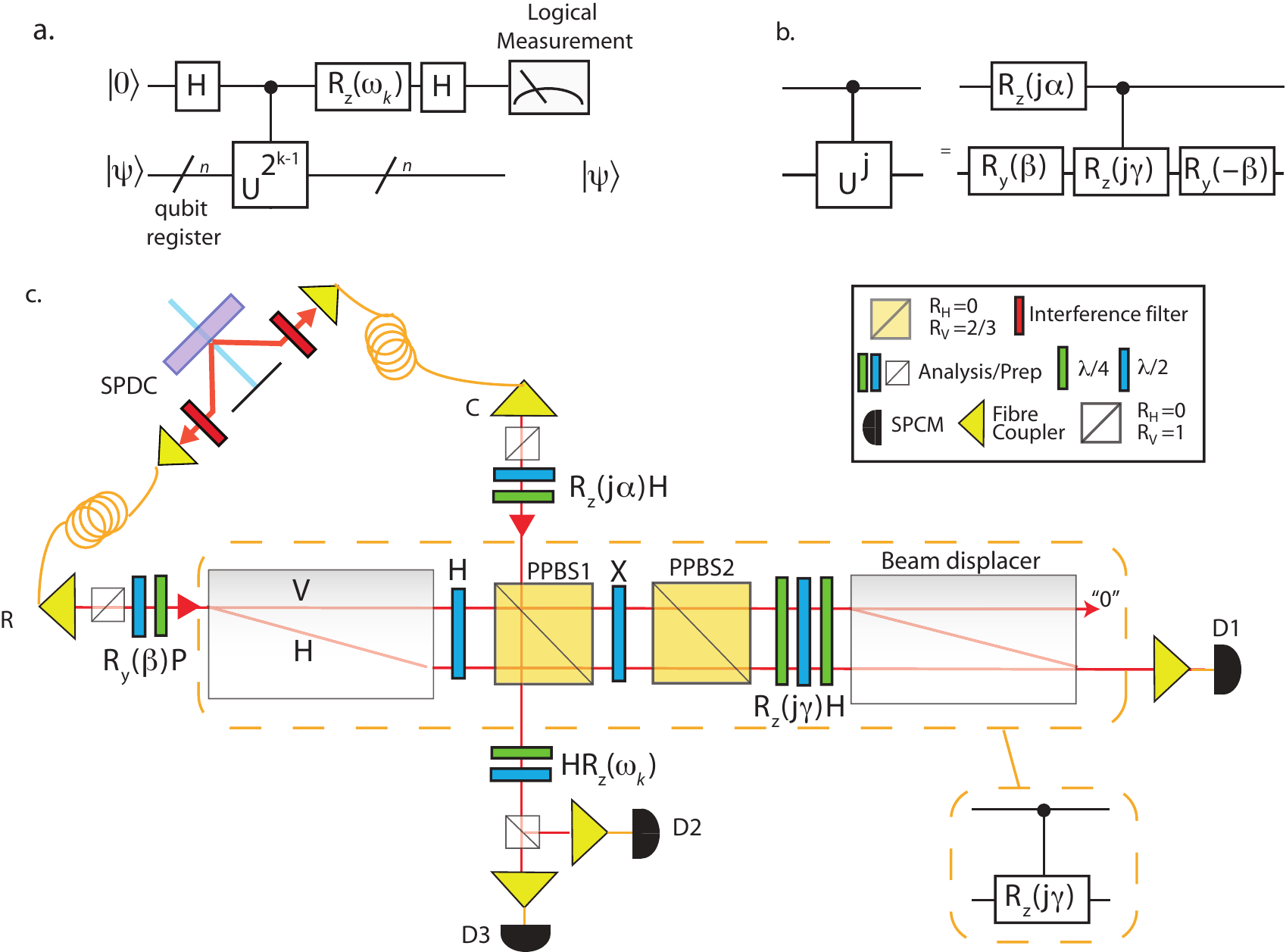}
\caption{\textbf{Algorithm and experimental implementation}.
\textbf{(a)} IPEA~\cite{AGDL+05,DJSW06} at iteration $k$. To produce a $K$-bit
approximation to $\phi$ the algorithm is iterated $K$ times. Each iteration obtains one
bit of $\phi$ ($\phi_k$): starting from the least significant ($\phi_K$), $k$ is iterated backwards from $K$ to 1. The angle $\omega_k$ depends on all previously measured bits, $\omega_k {=} -2\pi b$, where $b$, in the binary expansion, is $b{=}0.0\phi_{k{+}1}\phi_{k{+}2}...\phi_{K}$ and $\omega_K {=} 0$. H is the standard Hadamard gate~\cite{NC01}. 
\textbf{(b)}  Our gate network for a two-qubit controlled-$\hat{U}^j$ gate, as discussed in the Methods Summary. 
\textbf{(c)} Two-qubit optical implementation of (a). Photon pairs are generated by spontaneous parametric down-conversion (SPDC), coupled into single-mode optical fiber and launched into
free space optical modes C (control) and R (register). Transmission through a polarizing beamsplitter (PBS) prepares a photonic polarization qubit in the logical state $\ket{0}$, the horizontal polarization. The combination of a PBS with half ($\lambda/2$) and quarter ($\lambda/4$) waveplates allows the preparation (or analysis) of an arbitrary one-qubit pure state.  The optical controlled-$\hat{R}_z$ gate, shown in the dashed box, is realized using conditional transformations via spatial degrees of freedom as described by Lanyon~\cite{lanyon-2008} \emph{et al}. Coincident detection events (3.1 ns window) between single photon counting modules (SPCM's) D1 and D3 (D2 and D3) herald a successful run of the circuit and result 0 (1) for $\phi_k$. Waveplates are labelled with their corresponding operations.}\label{fig1}
\end{figure}

We implement the \emph{iterative} phase estimation algorithm~\cite{AGDL+05,DJSW06} (IPEA), which advantageously reduces the number of qubits and quantum logic gates required. Fig.\ref{fig1}~a shows the IPEA at iteration $k$. 
The result of a logical measurement of the top `control' qubit after each iteration determines the $k^{\mathrm{th}}$ bit of the binary expansion~\cite{NC01} of $\phi$. Let $K$ bits of this expansion be $\tilde{\phi}{=}0.\phi_1\phi_2...\phi_m$, such that $\phi{=}\tilde{\phi}{+}\delta2^{-K}$ where $\delta$ is a remainder $0 {\leq} \delta {<} 1$. An accuracy of $\pm 2^{-K}$ is achieved with error probability~\cite{DJSW06} $\epsilon {\le} 1{-}8/\pi^2 {\approx} 0.19$, which is independent of $K$ (the bound is saturated for $\delta{=}0.5$). This error can be eliminated by simply repeating each IPEA iteration multiple ($n$) times, yielding $n$ possible values for the corresponding bit; a majority vote of these samples determines the result (see Methods, Section C).

To resolve the energy differences relevant to chemical processes\cite{AGDL+05}, absolute molecular energies must be computed to an accuracy greater than ${\approx} 10^{-4}E_\mathrm{h}$ (${\sim} k_bT$ at room temperature). Therefore it is important to demonstrate that the IPEA can achieve the necessary phase precision of ${\approx} 16$ bits (the accuracy of the non-relativistic Born-Oppenheimer energy is then limited only by the choice of basis). We implement the IPEA with a photonic architecture, encoding qubits in polarization of single photons, Fig.\ref{fig1}~c. Our experiment is possible due to the recent development of a photonic two-qubit controlled-unitary quantum logic gate, which combines multiple photonic degrees of freedom, linear optical elements and projective measurement to achieve the required nonlinear interaction between photons\cite{lanyon-2008}. Such gates are high-quality, well-characterized, and have several in-principle paths to scalable optical quantum computing\cite{kok:135}. We note that our implementation of two-qubit quantum IPEA is the first, in any context, to use entangling gates, outside of a liquid-state ensemble NMR architecture\cite{XMJ+07}, which is arguably an in-principle non-scalable architecture\cite{PhysRevLett.83.1054}. We remark that an implementation of a semiclassical quantum Fourier transform was performed in ions\cite{Chia}, combining single-qubit measurement and rotations in place of entangling gates. 

\begin{figure}[p]
\includegraphics[width=.8 \columnwidth]{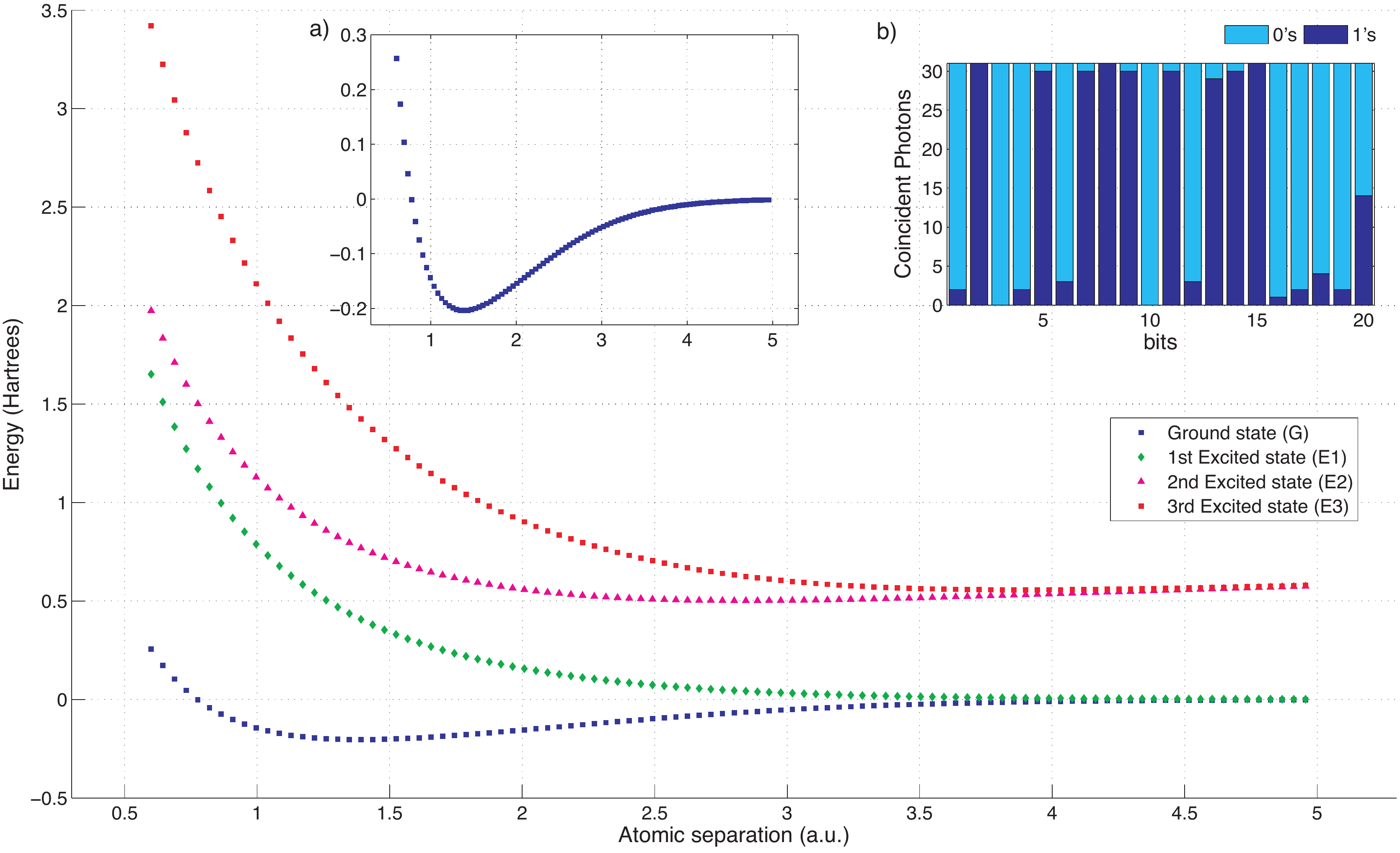}
\caption{\textbf{Quantum algorithm results: H$_2$ potential energy curves in a minimal basis}. Each point is calculated using a 20-bit IPEA and employing $n{=}31$ samples per bit (repetitions of each iteration). Every case was successful, achieving the target precision of $\pm (2^{-20} {\times} 2\pi)~E_{\textrm{h}} {\sim} 10^{-5}~E_{\textrm{h}}$.   Curve G (E3) is the low (high) eigenvalue of $\hat{H}^{(1,6)}$. Curve E1 is a triply degenerate spin-triplet state, corresponding to the lower eigenvalue of $\hat{H}^{(3,4)}$ as well as the eigenvalues $\hat{H}^{(2)}$ and $\hat{H}^{(5)}$. Curve E2 is the higher (singlet) eigenvalue of $\hat{H}^{(3,4)}$. Measured phases are converted to energies $E$ via $E{=}2\pi\phi {+}1/r$, where the last term accounts for the proton-proton Coulomb energy at atomic separation $r$, and reported relative to the ground state energy of two hydrogen atoms at infinite separation. 
\textbf{Inset a):} Curve G rescaled to highlight the bound state. \textbf{Inset b)}: Example of raw data for the ground state energy obtained at the equilibrium bond length, $1.3886~a.u.$. The measured binary phase is $\phi{=}0.01001011101011100000$ which is equal to the exact value, in our minimal basis, to a binary precision of ${\pm} 2^{-20}$. Note that the exact value has a remainder of $\delta {\approx} 0.5$ after a 20 bit expansion, hence the low contrast in the measured 20$^{th}$ bit.} \label{fig2}
\end{figure}

Fig.~\ref{fig2} shows our results: H$_2$ energies calculated over a range of internuclear separations, thus reconstructing the potential energy surfaces. Each point is obtained using a 20-bit IPEA and employing $n{=}31$ samples per bit. In every case, the algorithm successfully returned the energy to within the target precision of $\pm (2^{-20} {\times} 2\pi)E_\mathrm{h} {\approx} 10^{-5} E_\mathrm{h}$, in the minimal basis. For example, the ground state energy obtained at the equilibrium bond length, $1.3886~a_0$ (where $a_0$ is the Bohr radius), is $-0.20399 {\pm} 0.00001~E_\mathrm{h}$, which agrees exactly with the result obtained on a classical computer to an uncertainty in the least significant bit. 

Achieving such high precision will become a far more significant challenge for  large-scale implementations: due to the small-scale of our demonstration, we are able to implement each power of $\hat{U}^{(p,q)}$ directly, by re-encoding the same number of gates (see Methods Summary). Therefore, the probability of error introduced by gate imperfections remains a constant for each bit.  This is the main algorithmic feature that allows the high precision obtained in this experiment. However, following current proposals, the circuit network for $\hat{U}$ will not generally have the same form as $\hat{U}^j$ for larger implementations (detailed in the Appendix, Section A). For each additional digit of precision sought, the gate requirements of the algorithm are roughly doubled, thereby amplifying any gate error.

Important next experimental steps are to demonstrate the two parts of the quantum algorithm that we implemented with assistance from a classical computer. Firstly, encoding even low fidelity eigenstate approximations into qubits is a highly non-trivial step for molecules much larger than H$_2$. In many cases this problem could be overcome using a heuristic adiabatic state preparation algorithm~\cite{WBL02,AGDL+05, Friedenauer:2008qm}. Here, ground state approximations, for example, can be efficiently obtained provided that the energy gap between the ground state and excited states is sufficiently large along the path of the adiabatic evolution~\cite{FGGS00}. Secondly, directly calculating and decomposing the molecular evolution operator into logic gates does not scale efficiently with molecular size~\cite{Llo96} and an alternative scheme must be employed. The proposed solution exploits the fact that the general molecular Hamiltonian is a sum of fixed-sized one- and two-electron terms that can be individually efficiently simulated and combined to approximate the global evolution~\cite{Llo96, NC01}. 
We give an overview of this `operator-splitting technique' in the Appendix (Section A) and find that the total number of elementary quantum gates required to simulate the evolution of an arbitrary molecule, without error correction,  scales as $O(N^5)$, where $N$ is the number of single-particle basis functions used to describe the molecular system. In this scheme, $N$ is also the number of qubits necessary. We also present the quantum logic circuits required to simulate each term in the general molecular Hamiltonian---these are the building blocks of a universal quantum molecular simulator. Finally, we perform an accurate resource count to reproduce our H$_2$ simulation in this scalable way:  4 qubits and  ${\sim}522$ perfect gates are required to simulate the full unitary propagator such that the error of the simulated evolution is within chemical precision.

Other major challenges in the path to scalability include those associated with scaling up the `hardware', i.e., achieving more qubits, gates, and longer coherence times. Much progress is being made on developing the necessary technology for a large-scale photonic quantum computer~\cite{dowling, grangier}. The influence of noise is perhaps the most serious consideration~\cite{DBB07} and must be overcome using error-correction and fault-tolerant constructions~\cite{NC01,CBMG08}. 
We note that an alternative promising path to efficient quantum simulators is to exploit controllable quantum systems that can be used to \emph{directly} implement model Hamiltonians, thereby avoiding the aforementioned resource intensive approximation techniques and error correction\cite{Llo96, J03}.

\vspace{-3mm}
\section{Methods Summary}
\vspace{-3mm}
\noindent We use the minimal STO-3G basis~\cite{HSP69} for H$_2$, consisting of one $\ket{1s}$-type atomic orbital per atom. These two functions are combined to form the bonding (antibonding) molecular orbitals~\cite{SO96}. The 4 corresponding single-electron molecular spin-orbitals are combined antisymmetrically to form the six two-electron configurations ($\Phi_1 {\rightarrow} \Phi_6$) that form the basis for our simulation. Due to symmetry, the Hamiltonian is block-diagonal in this basis, with blocks acting on each of the four subspaces spanned by $\{\ket{\Phi_1},\ket{\Phi_6}\}$, $\{\ket{\Phi_2}\}$,$\{\ket{\Phi_3},\ket{\Phi_4}\}$, and $\{\ket{\Phi_5}\}$ (See Methods, Section A). Therefore, finding the eigenvalues of the two $2{\times}2$ sub-matrices in the Hamiltonian---$\hat{H}^{(1,6)}$ and $\hat{H}^{(3,4)}$---amounts to performing the FCI. Estimating the eigenvalues of 2x2 matrices is the simplest problem for the IPEA.

We employ a propagator time step of $t{=}1~\hbar/E_\mathrm{h}$ (the hartree, $E_\mathrm{h}{\approx}27.21$~eV, is the atomic unit of energy), chosen so that $0{\leq} Et/2\pi\hbar {\leq} 1$.  For our proof-of-principle demonstration, all necessary molecular integrals are evaluated classically (Methods, Section C) using the Hartree-Fock procedure~\cite{SO96}. We  use these integrals to calculate the matrix elements of $\hat{H}$ and $\hat{U}$, then directly decompose each $\hat{U}^{(p,q)}$ operator into a logic gate network. We decompose the $\hat{U}^{(p,q)}$ operators into a global phase and a series of rotations of the one-qubit Hilbert space~\cite{NC01}:
\begin{equation}
\hat{U}=e^{i\alpha}\hat{R}_y(\beta)\hat{R}_z(\gamma)\hat{R}_y(-\beta),
\label{eq2}
\end{equation}
where $\alpha$, $\beta$, and $\gamma$, are real angles. $\hat{U}^j$ is achieved by replacing angles $\alpha$ and $\gamma$ with $j\alpha$ and $j\gamma$ (while $\beta$ remains unchanged). Our decomposition of the controlled-$\hat{U}^j$ is shown in Fig.~\ref{fig1}b.\\

\noindent \textbf{Acknowledgments}.
\noindent We thank A. Perdomo, A. Steinberg, P.J. Love, A.D. Dutoi, G. Vidal, and A. Fedrizzi for discussions. We acknowledge financial support from the Australian Research Council (ARC) Federation Fellow and Centre of Excellence programs, and the IARPA-funded U.S. Army Research Office Contracts W911NF-0397 \& W911NF-07-0304. BJP was the recipient of an ARC Queen Elizabeth II Fellowship (DP0878523) and IK of the Joyce and Zlatko Balokovi{\'c} Scholarship.\\

\noindent \textbf{Correspondence}.
\noindent Correspondence and requests for materials should be addressed to BPL
(lanyon@physics.uq.edu.au) and AA-G (aspuru@chemistry.harvard.edu).\\

\vspace{-3mm}
\newpage
\section{Methods}
%\vspace{-5mm}
\subsection{Minimal basis and symmetries in the electronic Hamiltonian of the hydrogen molecule}
\vspace{-3mm}
\noindent The two $\ket{1s}$-type atomic orbitals are combined to form the bonding and antibonding molecular orbitals~\cite{SO96},
$\ket{g}$ and $\ket{u}$. Taking into account electron spin, the single-electron molecular spin-orbitals are denoted,
$\ket{g \! \uparrow}$,
$\ket{g \! \downarrow}$,
$\ket{u \! \uparrow}$ and
$\ket{u \! \downarrow}$,
where $\ket{\! \uparrow}$ and $\ket{\! \downarrow}$ are the electron spin
eigenstates. These are combined antisymmetrically to form the six
two-electron configurations that form the basis
for our simulation:
$\ket{\Phi_1}{\equiv}|g \! \uparrow, g \! \downarrow |{=}(\ket{g \! \uparrow, g \! \downarrow}{-}\ket{g \! \downarrow, g \! \uparrow})/\sqrt(2)$,
$\ket{\Phi_2}{=}|g \! \uparrow, u \! \uparrow |$,
$\ket{\Phi_3}{=}|g \! \uparrow, u \! \downarrow |$,
$\ket{\Phi_4}{=}|g \! \downarrow, u \! \uparrow |$,
$\ket{\Phi_5}{=}|g \! \downarrow, u \! \downarrow |$ and
$\ket{\Phi_6}{=}|u \! \uparrow, u \! \downarrow |$. 
Due to symmetry, the Hamiltonian is block-diagonal in this basis, with blocks acting on each of the four subspaces spanned by
$\{\ket{\Phi_1},\ket{\Phi_6}\}$, $\{\ket{\Phi_2}\}$,
$\{\ket{\Phi_3},\ket{\Phi_4}\}$, and $\{\ket{\Phi_5}\}$. 
Most of the elements of this basis are not mixed by the Hamiltonian. In particular, $\ket{\Phi_1}$ and $\ket{\Phi_6}$ mix only with each other because they have $g$ symmetry while the rest have $u$ symmetry.   Of the remaining states only $\ket{\Phi_3}$ and $\ket{\Phi_4}$ mix because they have the same total $z$-projection of the spin, $m_S{=}0$. $\ket{\Phi_2}$ and $\ket{\Phi_5}$ have, respectively, $m_S{=}1$ and $m_S{=}-1$. Therefore, the Hamiltonian is block-diagonal within four subspaces spanned by
$\{\ket{\Phi_1},\ket{\Phi_6}\}$, $\{\ket{\Phi_2}\}$, $\{\ket{\Phi_3},\ket{\Phi_4}\}$, and
$\{\ket{\Phi_5}\}$. There are no approximations involved here, and finding the eigenvalues of the two $2{\times}2$ sub-matrices in the Hamiltonian ($\hat{H}^{(1,6)}$ and $\hat{H}^{(3,4)}$) amounts
to performing an exact calculation (FCI) in the minimal basis. One should also note that it follows from the requirement that the wave functions are spin eigenstates, that the eigenstates of the subspace $\{\ket{\Phi_3},\ket{\Phi_4}\}$ will be $(\ket{\Phi_3} {\pm} \ket{\Phi_4})/\sqrt{2}$. Additionally, there will be a three-fold degeneracy of the triplet state with angular momentum $S{=}1$. That is, the states $\ket{\Phi_2}$, $\ket{\Phi_5}$, and $(\ket{\Phi_3}{+}\ket{\Phi_4})/\sqrt{2}$ are degenerate.

\vspace{-5mm}
\subsection{Details of computational methods}
\vspace{-3mm}
\noindent Restricted Hartree-Fock calculations were carried out on a classical computer using the STO-3G basis~\cite{HSP69}.  The software used was the PyQuante quantum chemistry package version 1.6.  The molecular integrals from the Hartree-Fock procedure are used to evaluate the matrix elements of the Hamiltonians $\hat{H}^{(1,6)}$ and $\hat{H}^{(3,4)}$, described in the main text.

\vspace{-5mm}
\noindent \subsection{Classical error correction technique}
\vspace{-3mm}
\noindent When running the IPEA, the probability of correctly identifying any individual bit with a single sample ($n{=}1$) is reduced from unity by both theoretical (inexact phase expansion to $K$ bits) and experimental factors (such as imperfect gates). However, as long as it remains above 0.5, repeated sampling and a majority vote will reduce the probability of error exponentially with $n$, in accordance with the Chernoff bound~\cite{NC01}. This technique allows for a significant increase in success probability, at the expense of repeating the experiment a fixed number of times. We note that this simple classical error correction technique can only play a small role when it comes to dealing with errors in large-scale implementations. Here, the numerous errors in very large quantum logic circuits will make achieving a bit success probability over  0.5 a significant challenge, that must be met with quantum error correction techniques~\cite{NC01,CBMG08}.\\

\subsection{Count rates}
\vspace{-3mm}
\noindent We operate with a low-brightness optical source (spontaneous parametric downconversion pumping power ${\approx} 50$~mW) to reduce the effects of unwanted multi-photon-pair emissions (which cannot be distinguished by our non-photon-number-resolving detectors and introduce error into the circuit operation). This yields about 15 coincident detection events per second at the output of our optical circuit. Therefore each iteration can be repeated 15 times a second. Reconfiguring the circuit for different iterations takes approximately 7 seconds, largely due to the finite time required to rotate standard computer controlled waveplate mounts. Therefore, obtaining a 20-bit estimation of a phase takes about 3 minutes, when using $n{=}31$ samples to determine the logical state of each bit (as was employed to achieve the results shown in Fig.~\ref{fig2}). Note that approximately 95\% of this time is spent rotating waveplates. In future implementations, this time could be reduced significantly using integrated-photonics, e.g. qubit manipulation using an electrooptically-controlled waveguide Mach-Zehnder interferometer~\cite{LXCZ+08}.\\

\newpage

\renewcommand{\theequation}{S\arabic{equation}}
\setcounter{equation}{0}
\setcounter{subsection}{0}
\newpage
\section{Appendix}

\subsection{Efficient simulation of arbitrary molecular time-evolution operators}

\noindent A fundamental challenge for the quantum simulation of
large molecules is the accurate decomposition of the system's
time evolution operator, $\hat{U}$. In our experimental demonstration, we
exploit the small size and inherent symmetries of the hydrogen molecule Hamiltonian to
implement $\hat{U}$ exactly, using only a small number of gates. As
the system size grows such a direct decomposition will no longer be practical. However, an efficient first-principles
simulation of the propagator is possible for larger chemical
systems~\cite{AGDL+05,AL99,Llo96,SOGK+02,OGKL01,OHJ07,VA08}.

The key steps of an efficient approach are: (1) expressing the
chemical Hamiltonian in second quantized form, (2) expressing each
term in the Hamiltonian in a spin $1/2$ representation via the
Jordan-Wigner transformation~\cite{JW28}, (3) decomposing the
overall unitary propagator, via a Trotter-Suzuki
expansion~\cite{HS05,Llo96}, into a product of the evolution
operators for non-commuting Hamiltonian terms, and (4) efficiently
simulating the evolution of each term by designing and implementing
the corresponding quantum circuit. We note that the first two steps
generate a Hamiltonian that can be easily mapped to the
state space of qubits. The last steps are part of the quantum algorithm
for simulating the time-evolution operator, $\hat{U}$, generated by
this Hamiltonian. Details of each step are provided as
follows:

\subsubsection*{Step 1. Second-quantized Hamiltonian}

\noindent The general second-quantized chemical
Hamiltonian has $O(N^4)$ terms, where $N$ is the number of single-electron basis
functions (i.e. spin-orbitals) used to describe the system~\cite{Helgaker2000}.   The Hamiltonian
can be explicitly written as:
\begin{equation}\label{eq:ham2nd}
    \hat{H}=\sum_{p,q}h_{pq}\ad_p \hat{a}_q +\frac{1}{2}\sum_{p,q,r,s} h_{pqrs} \ad_p \ad_q \hat{a}_r
    \hat{a}_s,
\end{equation}
where the annihilation and creation operators ($\hat{a}_j$ and $\ad_j$ respectively)
obey the fermionic anti-commutation relations: $[\hat{a}_i,\ad_j]_+=\delta_{ij}$ and $[\hat{a}_i,\hat{a}_j]_+=0$,
and the indices $p$, $q$, $r$, and $s$ run over all $N$ single-electron basis functions.
The integrals $h_{pq}$ and $h_{pqrs}$ are evaluated during a preliminary Hartree-Fock procedure~\cite{SO96} and are defined as
\begin{eqnarray}
    h_{pq}&=&\int\d\mathbf{x} \;\chi^*_p(\mathbf{x})
        \left(-\frac{1}{2}\nabla^2-\sum_\alpha \frac{Z_\alpha}{r_{\alpha\mathbf{x}}}\right)
                     \chi_q(\mathbf{x})\nonumber
\end{eqnarray}
and
\begin{eqnarray}
h_{pqrs}&=&  \int \d\mathbf{x}_{1} \d\mathbf{x}_{2}
    \frac{ \chi^*_p(\mathbf{x}_{1})\chi_q^*(\mathbf{x}_{2})\chi_r(\mathbf{x}_{2})\chi_s(\mathbf{x}_{1})}{r_{12}}
    \nonumber
\label{eq:2eint}
\end{eqnarray}
where $\chi_q(\mathbf{x})$ are a selected single-particle basis. Here
$\nabla^2$ is the Laplacian with respect to the electron spatial
coordinates, while $r_{\alpha\mathbf{x}}$ and
$r_{12}$ are the distances between the
$\alpha^\textrm{th}$ nucleus and the electron and the
distance between electrons 1 and 2, respectively.

Expressing the Hamiltonian in second-quantized form allows straightforward
mapping of the state space to qubits. The logical states of each qubit are
identified with the fermionic occupancy of a single-electron spin-orbital (i.e.
$\ket{0}=$~occupied, $\ket{1}=$~unoccupied). Therefore, simulating a system with
a total of $N$ single-electron spin-orbitals (e.g., $N=\lambda\kappa$ for a
molecule with $\lambda$ atoms each with $\kappa$ spin-orbitals) requires only $N$
qubits. Note that the $N$-qubit Hilbert space allows for any number of electrons
(up to $N$), hence the scaling is \emph{independent} of the number of electrons
present in the system. In practical Gaussian basis-set calculations, the number
of spin-orbitals per atom is usually constant for a given row of the periodic
table~\cite{SDES+07}. The use of a double-zeta basis set~\cite{SDES+07} would require employing $\approx$ 30 logical qubits per simulated atom. For example, 1800 logical qubits would be required to store the wave function of the fullerene (C$_{60}$) molecule.

\subsubsection*{Step 2. Jordan-Wigner transformation of the fermionic operators to spin variables}

\noindent Starting with the second-quantized Hamiltonian from~\eqref{eq:ham2nd}, the Jordan-Wigner transformation is used to map fermionic creation and annihilation operators into a representation using the Pauli spin matrices as a basis~\cite{JW28}. This allows
for a convenient implementation on a quantum
computer~\cite{SOGK+02,OGKL01}. The representation is achieved via the following invertible
transformations, which are applied to each term in~\eqref{eq:ham2nd}:
\begin{subequations}\label{:JW}
\begin{eqnarray}
    \hat{a}_j& \rightarrow&{\I}^{\otimes j-1}\otimes \hat{\sigma}^{+} \otimes \left(\hat{\sigma}^{z}\right)^{\otimes N-j}\label{subeq:JW(dest)}\\
    \ad_j&\rightarrow&{\I}^{\otimes j-1}\otimes { \hat{\sigma}^{-}} \otimes \left(\hat{\sigma}^z\right)^{\otimes N-j},\label{subeq:JW(crea)}
\end{eqnarray}
\end{subequations}
where $\hat{\sigma}^+\equiv (\hat{\sigma}^x+i\hat{\sigma}^y)/2=\ket{0}\bra{1}$
and $\hat{\sigma}^-\equiv ({\hat{\sigma}^x-i\hat{\sigma}^y})/2=\ket{1}\bra{0}$.  The $\hat{\sigma}^\pm$ operators achieve
the desired mapping of occupied (unoccupied) states to the computational basis
[i.e., $|1\rangle$ ($|0\rangle$)] while other terms serve to
maintain the required anti-symmetrization of the wavefunction in the
spin (qubit) representation.

\subsubsection*{Step 3. Exponentiation of the Hamiltonian}

As the system size represented by the chemical Hamiltonian \eqref{eq:ham2nd} grows, a direct decomposition of the time-evolution operator, $\hat{U}$, into a sequence of logic gates will no longer be practical as the best methods scale exponentially.   However, the Hamiltonian is a sum of one and two-electron terms whose time-evolution operators can each be implemented efficiently---e.g. with a number of gates that does not scale with $N$. However, generally the terms do not commute, thus simple reconstruction of $\hat{U}$ from direct products of the individual operators is not possible. Trotter-Suzuki relations can be used to approximate the full unitary propagator from the individual evolution of non-commuting operators~\cite{HS05,Llo96}. 

For a Hamiltonian $\hat{H}=\sum_{i=1}^N \hat{h}_i$,
the first-order Trotter-Suzuki decomposition is expressed as
\begin{equation}
\hat{U}(t)=e^{-i\hat{H}t}= \left(e^{-i\hat{h}_1dt}e^{-i\hat{h}_2 dt}\cdots
e^{-i\hat{h}_Ndt}\right)^{\frac{t}{dt}}+O(dt^2).  \label{eq:Trotter1stOrder}
\end{equation}
The value $T_n=t/dt$ is called the Trotter number~\cite{HS05}.  As the Trotter number tends to
infinity, or equivalently $dt\rightarrow 0$, the approximation
becomes exact. In practice, a compromise between computational effort and accuracy is employed. In
numerical computations, 
successive calculations at different timesteps $dt$
are often carried out, and an extrapolation of $dt\to 0$ gives an estimate
of the exact answer. A similar approach can be used for quantum
simulation.

We note that, unlike our small-scale experiment, the powers of the system
evolution operator, $\hat{U}^j$, required for the IPEA cannot be achieved by
simply changing parameters in the gate decomposition for $\hat{U}$. In general
$\hat{U}^2$ will take twice as many gates as $\hat{U}$. Intuitively, the system
dynamics must be propagated for twice as long leading to twice as many
manipulations of the quantum simulator's natural dynamics. The increase in the
number of gates required for extra bits will clearly amplify experimental
errors, thereby limiting the obtainable precision. Note that although the number
of required gates increases exponentially with the number of bits, each
additional bit itself provides an exponential increase in precision.

As mentioned in the manuscript, quantum algorithms that circumvent the problems found from the Trotter expansion are a fertile area of research. In the current scheme, Hamiltonians that are diagonal in the computational basis, such as the classical Ising model do not require a Trotter expansion for their accurate simulation~\cite{CBMG08}.

\subsubsection*{Step 4. Circuit representations of the unitary propagator}

\noindent Each exponentiated tensor product of Pauli matrices can then be
implemented efficiently by employing a family of quantum circuits. In order to
provide an accurate estimation of an upper bound of the number of gates required
for the different kinds of second-quantized operators, we carried out analytical
gate decompositions. The circuit networks obtained are summarized in
Fig.~\ref{fig:TrotterTable}. The networks shown realize the unitary operator
$\hat{U}(dt)$ for a general molecular Hamiltonian. To realize a controlled
unitary, $c-\hat{U}(dt)$, as required by the phase estimation algorithm, only
the rotations $\hat{R}_z(\theta)$ must be converted to
controlled-$\hat{R}_z(\theta)$ rotations. The number of gates required to
simulate each term is linear in the number of intervening qubits due to the
product of $\hat{\sigma}_z$ terms resulting from the Jordan-Wigner
transformation of Eq. \ref{:JW}. Therefore, the scaling of the number of quantum
gates required for simulating a general many-electron chemical Hamiltonian is
$O(N^5)$ without considering the influence of noise~\cite{DBB07}. Fault tolerant
quantum simulation~\cite{NC01} requires the use of a finite set of gates and the
conversion from the continuous set of gates to a discrete set can be
accomplished with polylogarithmic overhead~\cite{Kit97}. The encoding of robust
quantum states will also require several redundant qubits for each logical qubit
needed~\cite{NC01}. A more detailed analysis of fault tolerance in the context
of quantum simulation can be found in Ref.~\cite{CBMG08}.

\subsubsection*{Resource count for a simple example.}

\noindent In order to illustrate this algorithm, we performed numerical
simulations for H$_2$ in the same minimal basis (STO-3G) employed in our
experiment. Unlike our experimental mapping, the logical states of each register
qubit are now identified with the fermionic occupancy of the four
single-electron spin-orbitals (i.e. $\ket{0}=$~occupied, $\ket{1}=$~unoccupied).
Therefore, the calculation requires a total of five qubits taking into
consideration the single control qubit required for the IPEA. If quantum error
correction is needed, the number of qubits will increase according to the scheme
used~\cite{NC01}. Fig.~\ref{fig:H2} shows the error in the ground state energy
as a function of the Trotter step. The ground state energies of the approximate
unitary propagators were obtained via direct diagonalization on a classical
computer. A precision of $\pm 10^{-4}E_h$ is achieved at a Trotter number of 6,
which corresponds to 522 gates. Note that this gate count is to construct
$\hat{U}^1$ and includes both one- and two-qubit operations. This estimate does
not take into consideration error correction for the qubits and it uses a
continuous set of gates. In the path to large scale implementations, both will
be serious considerations and will increase the complexity of the algorithm and
the number of qubits necessary~\cite{NC01,CBMG08}. The unitary matrix must be
raised to various powers to perform phase estimation. If one desires to maintain
a fixed accuracy of 13 bits, about $8.5\times10^6$ gates must be used for the
IPEA estimation procedure. Note that this can be achieved by repeating the 522
gates required for $\hat{U}$ many times. Note that this does not include the
resources associated with preparing a system eigenstate. If one uses an
adiabatic state preparation techniques~\cite{AGDL+05} the resources are
proportional to the gap between the ground state and the excited state along the
path of adiabatic evolution~\cite{FJSM00}.

Although the estimates just given exceed the capabilities of current quantum
computers, these resource requirements grow only \emph{polynomially} with the
size the of system. Consequently, for large enough chemical systems, quantum
computers with around 100 qubits are predicted to outperform classical
computational devices for the first-principles calculation of chemical
properties \cite{AGDL+05,KJLM+08}.

\subsection{Additional experimental results}

\noindent We use the estimation of the ground state energy at the equilibrium bond length, $1.3886~a_0$ (where $a_0$ is the Bohr radius) to study the effect of varying a range of experimental parameters on the IPEA success probability, where we define the latter as the probability of correctly obtaining the phase to an precision of $2^{-m}$. Fig.~\ref{figs3}a shows results measured over a range of $n$, the number of samples used to determine each bit.

The probability of correctly identifying any individual bit with a single sample ($n=1$) is reduced from unity by both theoretical ($\delta$) and experimental factors (such as imperfect gates). However, as long as it remains above 0.5, repeated sampling and a majority vote will improve the probability of correct identification. The data show that this is achieved and the error
probability decreases exponentially with $n$, in accordance with the Chernoff bound~\cite{NC01}. This technique allows for a significant increase in success probability, at the expense of repeating the experiment a fixed number of times. We note that this simple classical error correction technique can only play a small role when it comes to dealing with errors in large-scale implementations. Here, the numerous errors in very large quantum logic circuits will make achieving a bit success probability over  0.5 a significant challenge, that must be met with quantum error correction techniques~\cite{NC01,CBMG08}.

Fig.~\ref{figs3}b shows the algorithm success probability measured as a function of the number of
extracted bits (phase precision). By employing $n=101$ samples per bit we achieve near perfect
algorithm success probability up to 47 bits (yielding an energy precision of
$\approx 10^{-13}E_\mathrm{h}$), where this limit is imposed only by the
machine-level precision used for the classical preprocessing of the
Hamiltonians. It is insightful to understand how achieving such high precision
will become a far more significant challenge for  large-scale implementations:
due to the small-scale of our demonstration, we are able to implement each power
of $\hat{U}^{(i,j)}$ directly, by re-encoding the same number of gates.
Therefore, the probability of error introduced by gate imperfections remains a
constant for each bit (and, in our implementation, under 50\%). This is the main
algorithmic feature that allows the high precision obtained in this experiment.
However, as expounded in the appendix (section A), this will not be possible for
larger implementations. In general, $\hat{U}$ will not have the same form as
$\hat{U}^n$. For each additional digit of precision sought, the gate
requirements of the algorithm are roughly doubled, thereby amplifying any gate error.

Fig.~\ref{figs3}c shows the algorithm success probability measured as a function of the fidelity $F$ (see caption) between the encoded register state and the ground state. The results show that our implementation is robust for $F \gtrsim 0.5$. Because the probability of correctly obtaining each bit in a single measurement ($n=1$) is greater than 0.5 in this regime, multiple sampling ($n>1$) enables the success probability to be amplified arbitrarily close to unity. This is a general feature that will hold for large-scale implementations. However, for $F\lesssim 0.5$, the measured success probabilities are very low.

If the register state output after each iteration is used as the input of the next, then the problem with low eigenstate fidelities can be overcome as the measurement of the control qubit collapses the wave function.  Any pure encoded register state can be written in the eigenstate basis as $\ket{G}=\sum_i\alpha_i\ket{\lambda_i}$, where $|\alpha_i|^2$ is the fidelity of $\ket{G}$ with eigenstate $\ket{\lambda_i}$. Successful measurement of the $m^{\mathrm{th}}$ bit associated with $\ket{\lambda_i}$ will cause the register wavefunction to collapse into a state with a greater fidelity with $\ket{\lambda_i}$---those eigenstates with a low probability of returning the measured bit value will be diminished from the superposition. As more bits are successfully measured, the register state will rapidly collapse to $\ket{\lambda_i}$.
In this way, the algorithm will return all the bits associated with $\ket{\lambda_i}$ with probability at least~\cite{NC01} $|\alpha_i|^2(1-\epsilon)$.
With current technology, correct operation of our optical circuit requires destructive measurement of both the control and register qubits after each IPEA iteration. Therefore, in our experiment the register state must be re-prepared for each iteration.

\subsubsection{How we obtain IPEA success probabilities}

\noindent Denoting the first $m$ binary bits of a phase $\phi$ as
$\tilde{\phi}=0.\phi_1\phi_2...\phi_m$, there is, in general, a remainder $0\leq
\delta < 1$, such that $\phi=\tilde{\phi}+\delta2^{-m}$. To achieve an accuracy
of $\pm{2^{-m}}$ the IPEA success probability is the sum of the probabilities
for obtaining $\tilde{\phi}$ and $\tilde{\phi}+2^{-m}$. This can be estimated
experimentally, for a given phase, by simply repeating the algorithm a large
number of times and dividing the number of acceptable results by the total. An
estimate with an error less than 10\% would require over 100 algorithm
repetitions. We calculate the result shown in Fig.~\ref{figs3}c in this way. However,
using this technique to obtain Fig.~\ref{figs3}b-c, and Fig.~\ref{figs3} (described below), would
take a long time---the 20 points shown in each would require more than 100 hours
of waveplate rotation time alone. Instead, to obtain these results we force the
appropriate feedforward trajectory ($R(\omega_k)$) for each accepted phase value
and use $n=301$ samples to estimate the 0/1 probabilities for each bit. Using
the standard binomial cumulative distribution function it is then possible to calculate
the majority vote success probability for each bit of each accepted value for a
given $n$ (1 and 101 in the figures). The probability for obtaining an accepted
phase value is then the product of the majority vote success probabilities for
each bit, and the total algorithm success probability is the sum of the
probabilities for obtaining each accepted phase. The error bars represent a 68\%
confidence interval and are obtained from a direct Monte-Carlo simulation of the
above process.

Note that forcing the correct feedforward in this way, and taking many samples
to estimate the 0/1 probabilities for each bit, simply allows us to accurately
estimate the probability that the algorithm will return the correct phase by
itself - i.e. without forcing the correct feedforward.

\subsubsection{Experimental model}

\noindent A simple computational model of our experiment produced the lines shown in Fig.~\ref{figs3}. This model allows for two experimental imperfections, which are described below, but otherwise assumes perfect optic element operation. The
model consists of a series of operators, representing optical elements and noise
sources, acting on a vector space representing both photonic polarisation and
longitudinal spatial mode~\cite{lanyon-2008}. Firstly the model allows for
photon distinguishability, quantified by an imperfect relative non-classical
interference visibility of 0.93 (ideal 1), which reduces the quality of our
two-qubit logic gate. Secondly the model allows for phase damping of the control
qubit, described by the operation elements~\cite{NC01}:

\begin{equation}
\left[
\begin{matrix}
1 & 0 \\
0 & \sqrt{1-\gamma}
\end{matrix}
\right]\;\; \text{and} \;\;
\left[
\begin{matrix}
0 & 0 \\
0 & \sqrt{\gamma}
\end{matrix}
\right].
\label{eqn:1}
\end{equation}

\noindent Our model employs $\gamma=0.06$ (ideal 0), which corresponds to $\approx 3\%$ dephasing. These experimental imperfections are attributed to a combination of residual higher-order photon pair emissions from our optical source and circuit alignment drift during long measurement sets.

\newpage


\renewcommand{\thefigure}{S\arabic{figure}}
\setcounter{figure}{0}

\begin{figure}[p]
\includegraphics[width=.7\columnwidth]{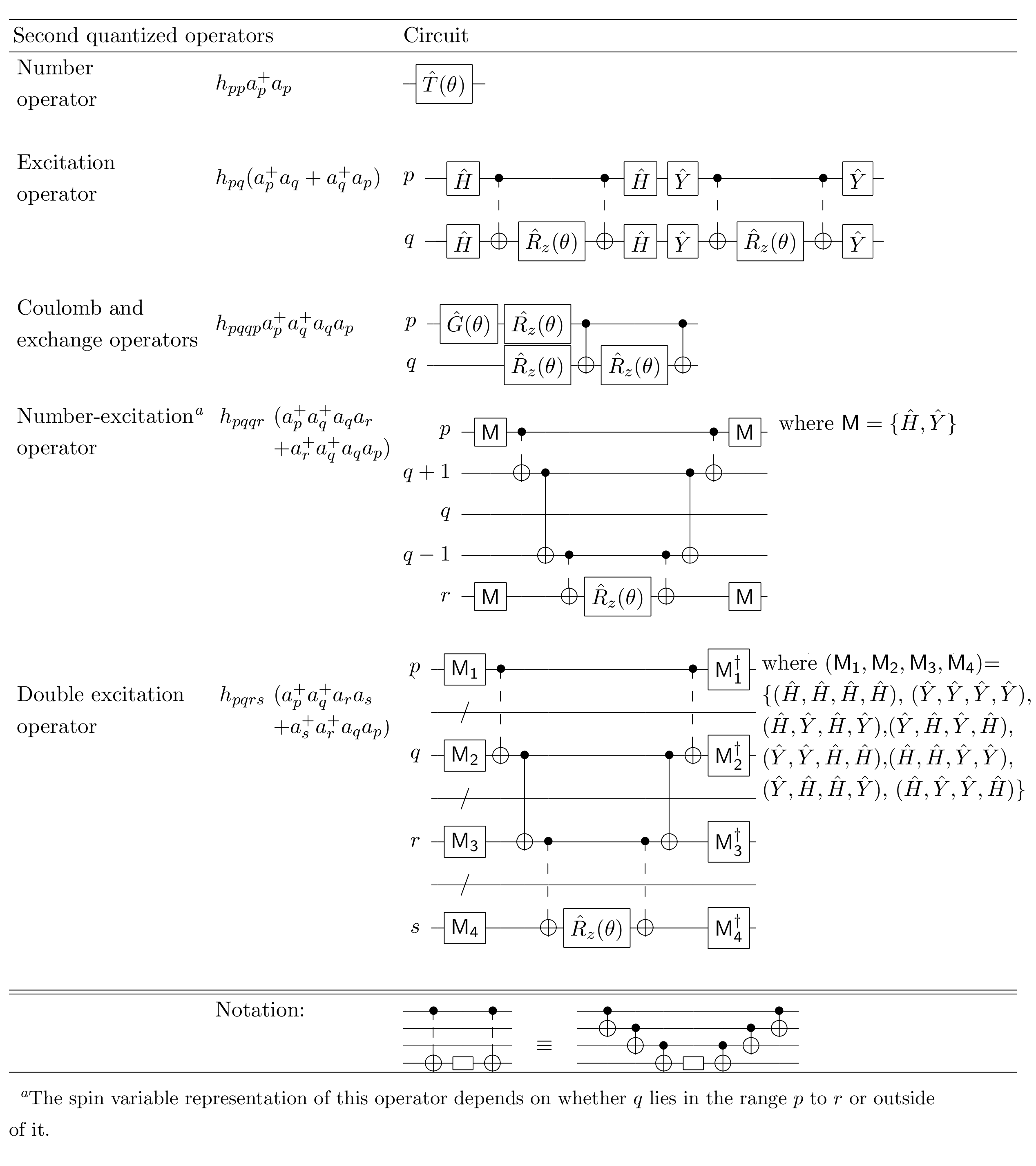}
    \caption{
    \textbf{The quantum circuits corresponding to evolution of the listed Hermitian second-quantized operators}. Here $p$, $q$, $r$, and $s$ are orbital indices corresponding to qubits such that the population of $\ket{1}$ determines the  occupancy of the orbitals.
 It is assumed that the orbital indices satisfy $p>q>r>s$. These circuits were found by performing the Jordan-Wigner transformation given in \eqref{subeq:JW(crea)} and \eqref{subeq:JW(dest)} and then propagating the obtained Pauli spin variables~\cite{SOGK+02}. In each circuit, $\theta=\theta(h)$ where $h$ is the integral preceding the operator.  Gate $\hat{T}(\theta)$ is defined by $\hat{T}\ket{0}=\ket{0}$ and $\hat{T}\ket{1}=\exp(-i\theta)\ket{1}$, $\hat{G}$ is the global phase gate given by $\exp(-i\phi)\hat{1}$, and the change-of-basis gate $\hat{Y}$ is defined as $\hat{R}_x(-\pi/2)$. Gate $\hat{H}$ refers to the Hadamard gate.  For the number-excitation operator, both ${\sf M}=\hat{Y}$ and ${\sf M}=\hat{H}$ must be implemented in succession. Similarly, for the double excitation operator each of the 8 quadruplets must be implemented in succession.   The global phase gate must be included due to the phase-estimation procedure.  Phase estimation requires controlled versions of these operators which can be accomplished by changing all gates with $\theta$-dependence into controlled gates. 
    }
\label{fig:TrotterTable}
\end{figure}

\begin{figure}[p]
    \includegraphics[width=1\columnwidth]{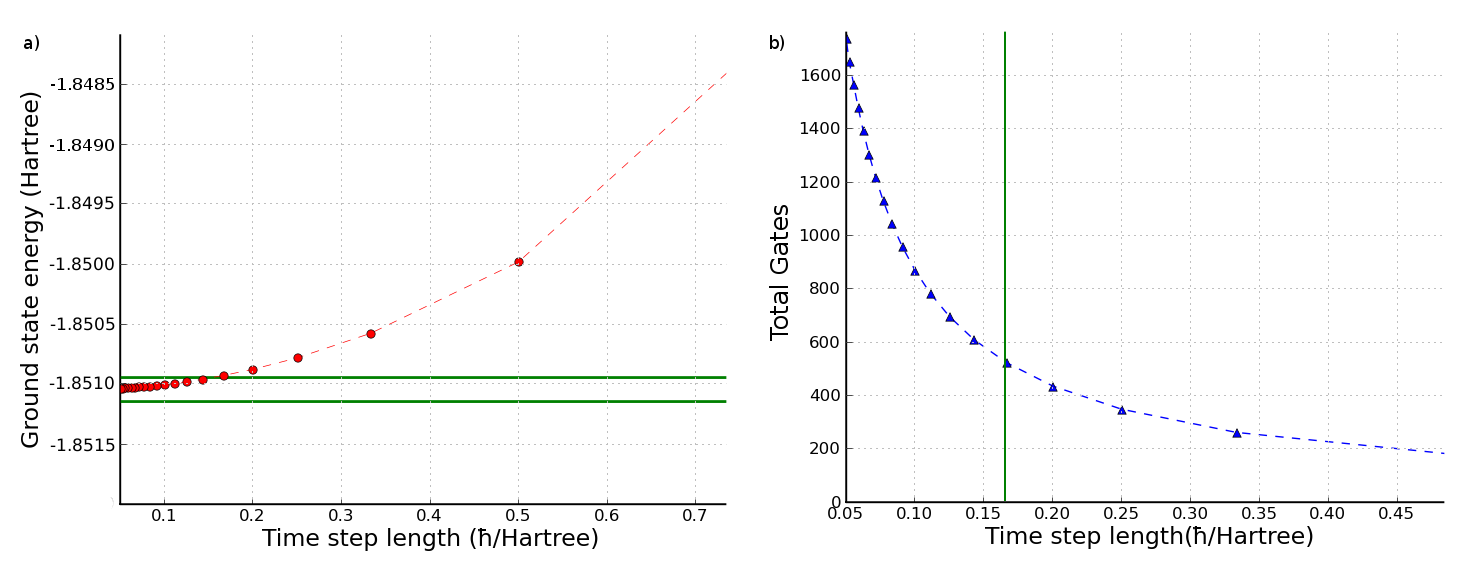}
    \caption{\textbf{Trotter error analysis and resource count for hydrogen molecule using a scalable quantum simulation algorithm}. \textbf{(a)} Plot of ground state energy of hydrogen molecule as a function of the length of the time step. As the time step length decreases, the accuracy of the approximation increases in accordance with eqn.~\eqref{eq:Trotter1stOrder}.  The total time of propagation, $t$, was unity and this time was split into time steps, $dt$.  The circles are at integer values of the Trotter number, $T_n\equiv t/dt$.  Green horizontal lines indicate the bounds for $\pm10^{-4}E_h$ precision.    \textbf{(b)} Gates for a single construction of the approximate unitary as a function of time step. As the time step decreases, more gates must be used to construct the propagator.  The triangles indicate integer values of the Trotter number and the green vertical line corresponds to the same threshold from graph a.  Perfect gate operations are assumed.
}
    \label{fig:H2}
\end{figure}

\begin{figure}[p]
\includegraphics[width=0.4 \columnwidth]{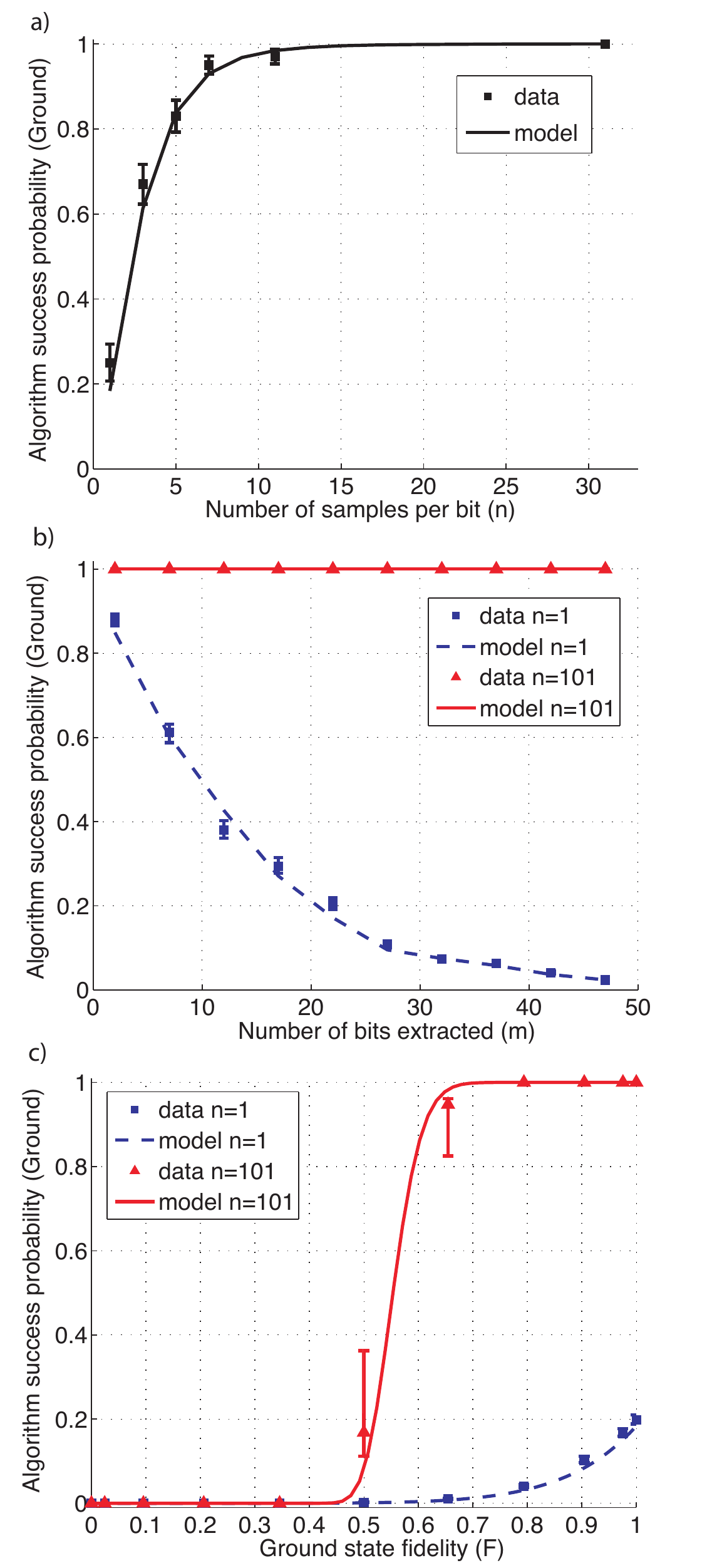}
\caption{\textbf{IPEA success probability measured over a range of
parameters.}
Probabilities for obtaining the ground state energy, at the equilibrium bond length $1.3886~a_0$, as a function of:
\textbf{(a)}  the number of times each bit is sampled ($n$);
\textbf{(b)} the number of extracted bits (m);
\textbf{(c)} the fidelity between the encoded register state and the ground state ($F$). 
The standard fidelity~\cite{NC01} between a measured mixed $\rho$ and ideal pure $\ket{\Psi}$ state is $F{=}\langle\Psi|\rho|\Psi\rangle$. 
\textbf{(a)} \& \textbf{(b)} employ a
ground state fidelity of $F\approx 1$. \textbf{(a)} \& \textbf{(c)} employ a 20-bit IPEA.
All lines are calculated
using a model that allows for experimental imperfections. This model, as well as the
technique used to calculate success probabilities and error bars, are detailed in the appendix (section B).}
\label{figs3}
\end{figure}

\end{document}